%% file: paper.tex
\documentclass[ussrhead]{elsart}
\usepackage{epsfig}
\usepackage{amssymb}

\def\hc#1{\hbox to\hsize{\hss #1\hss}}

\journal{{\tt Astroparticle Physics}}

\runauthor{T.~Antoni et al. (KASCADE Collaboration) }
\runtitle{Pure Beams}

\begin{document}

\begin{frontmatter}

\title{Preparation of enriched cosmic ray mass groups with KASCADE}

\author[KA-Uni]{T.~Antoni},
\author[KA-FZK]{W.\,D.~Apel},
\author[KA-Uni]{A.F.~Badea\thanksref{leaveBu}},\relax 
   \thanks[leaveBu]{on leave of absence from the NIPNE, 
   Bucharest, Romania}
\author[KA-FZK]{K.~Bekk},
\author[KA-FZK]{A.~Bercuci},
\author[KA-FZK,KA-Uni]{H.~Bl\"umer},
\author[BU]{H.~Bozdog},
\author[BU]{I.\,M.~Brancus},
\author[KA-Uni]{C.~B\"uttner},
\author[YE]{A.~Chilingarian},
\author[KA-Uni]{K.~Daumiller},
\author[KA-FZK]{P.~Doll},
\author[KA-FZK]{J.~Engler},
\author[KA-FZK]{F.~Fe{\ss}ler},
\author[KA-FZK]{H.\,J.~Gils},
\author[KA-Uni]{R.~Glasstetter},
\author[KA-Uni]{R.~Haeusler},
\author[KA-FZK]{A.~Haungs},
\author[KA-FZK]{D.~Heck},
\author[KA-Uni]{J.\,R.~H\"orandel},
\author[KA-Uni]{A.~Iwan\thanksref{leaveLo}},\relax 
   \thanks[leaveLo]{also University of Lodz, 
                    Lodz, Poland}
\author[KA-Uni,KA-FZK]{K.-H.~Kampert},
\author[KA-FZK]{H.\,O.~Klages},
\author[KA-FZK]{G.~Maier},
\author[KA-FZK]{H.-J.~Mathes},
\author[KA-FZK]{H.\,J.~Mayer},
\author[KA-Uni]{J.~Milke},
\author[KA-FZK]{M.~M\"uller},
\author[KA-FZK]{R.~Obenland},
\author[KA-FZK]{J.~Oehlschl\"ager},
\author[KA-Uni]{S.~Ostapchenko\thanksref{leaveMo}},\relax 
   \thanks[leaveMo]{on leave of absence from Moscow State University, 
                    Moscow, Russia}
\author[BU]{M.~Petcu},
\author[KA-FZK]{H.~Rebel},
\author[KA-FZK]{M.~Risse},
\author[KA-FZK]{M.~Roth},
\author[KA-FZK]{G.~Schatz},
\author[KA-FZK]{H.~Schieler},
\author[KA-FZK]{J.~Scholz},
\author[KA-FZK]{T.~Thouw},
\author[KA-Uni]{H.~Ulrich},
\author[YE]{A.~Vardanyan\thanksref{corres}},\relax 
   \thanks[corres]{corresponding author; email: aro@crdlx5.yerphi.am}%
\author[KA-Uni]{J.\,H.~Weber},
\author[KA-FZK]{A.~Weindl},
\author[KA-FZK]{J.~Wentz},
\author[KA-FZK]{J.~Wochele},
\author[LZ-Sol]{J.~Zabierowski}

\collab{(The KASCADE Collaboration)}

\address[KA-Uni]{Institut f\"ur Experimentelle Kernphysik, University of
             Karlsruhe, 76021~Karlsruhe, Germany}
\address[KA-FZK]{Institut f\"ur Kernphysik, Forschungszentrum Karlsruhe,
      	     76021~Karlsruhe, Germany}
\address[BU]{National Institute of Physics and Nuclear Engineering, 
             7690~Bucharest, Romania}
\address[YE]{Cosmic Ray Division, Yerevan Physics Institute, 
             Yerevan~36, Armenia}
\address[LZ-Sol]{Soltan Institute for Nuclear Studies,
             90950~Lodz, Poland}
	     
\ifx AA
\makeatletter
\begingroup
  \global\newcount\c@sv@footnote
  \global\c@sv@footnote=\c@footnote     
  \output@glob@notes  
  \global\c@footnote=\c@sv@footnote     
  \global\t@glob@notes={}
\endgroup
\makeatother
\fi

\newpage

\begin{abstract}
\noindent The KASCADE experiment measures a high number of EAS 
observables with a large degree of sampling of the
electron-photon, muon, and hadron components.
It provides accurate data for an
event-by-event analysis of the primary cosmic ray flux in 
the energy range around the knee. 
The possibility of selecting samples of enriched proton and iron induced 
extensive air showers by applying the statistical techniques 
of multivariate analyses is scrutinized using detailed Monte Carlo
simulations of three different primaries.
The purity and efficiency of the proton and iron classified events 
is investigated.
After obtaining enriched samples from the measured data 
by application of the
procedures the reconstructed
number of hadrons, hadronic energy and other parameters are 
investigated in the primary energy range $10^{15}-10^{16}\,$eV.
By comparing these shower parameters for purified proton 
and iron events, respectively, with simulated distributions 
an attempt is made to check the validity of
strong interaction models at high energies.
\end{abstract}


\end{frontmatter}

\input{papertxt}

\bibliographystyle{arounsrt}
\newcommand{\noopsort}[1]{} \newcommand{\printfirst}[2]{#1}
  \newcommand{\singleletter}[1]{#1} \newcommand{\switchargs}[2]{#2#1}

\end{document}

%% file: papertxt.tex
%
%
%
\section{Introduction}\label{mech}

Above primary energies of a few hundred TeV direct measurements 
of energy and mass of individual cosmic ray nuclei are unfeasible due 
to the drastic decrease of the cosmic ray intensity with increasing 
energy.
Hence, one has resort to the measurements of extensive air showers 
(EAS) which are produced when high energy cosmic ray particles 
enter into the Earth's atmosphere. 
Therefore the determination of primary energy and mass from EAS 
observables  
depends on the understanding of the high-energy hadronic
interaction features of the primary particle, 
and further on of the shower development in general. 
Consequently redundant information on the measurements is 
required to disentangle the problem. 

The idea to use advanced statistical techniques of multivariate 
analyses~\cite{ch-stat-dec-cpc} for enriching certain classes of  
primaries~\cite{ch-strong-inter,ch-compos-cimento} and to prepare 
enriched samples by event-by-event analyses of 
EAS observations was first investigated for the ANI –
experiment~\cite{ch-ani-exp}.
The realization has become feasible
by recent measurements of the multi-detector –
experiment KASCADE~\cite{kascade1} which provides an  
accurate experimental basis by simultaneous measurements of 
many EAS observables for each individual event. 
The purpose of this paper is to apply these techniques to
KASCADE data and to investigate possibilities of testing 
high-energy hadronic interaction models.

Such an approach appears to be very promising in view of 
detailed tests of interaction models currently under debate
and for paving the way 
to a consistent description of the hadronic 
interaction at extremely high energies by experimental signatures. 
Still the results of the KASCADE  experiment concerning the 
energy spectrum and mass composition of primary 
cosmic rays  are considerably affected by the uncertainties of the
used Monte Carlo models, which are estimated to be 
much larger than the 
statistical uncertainties, e.g. for the deduced features 
of the cosmic ray energy spectrum~\cite{roth01}. 

The present investigation introduces the preparation of 
samples of enriched cosmic ray mass groups 
and their use for studies of hadronic
interactions with air nuclei. The concept for the
classification is based on multivariate nonparametric methods of 
statistical inference. Using the information on an event-by-event
basis empirical statements can be drawn on the validity of the 
`a priori' knowledge of the Monte Carlo simulations.    
Global event observables
like the muon and electron shower sizes are used in order to select 
event samples with enriched contents of proton and iron primaries,
respectively. Additional observables mainly of the hadronic 
component are subsequently used for the investigation of 
interaction features of the primaries.  
It is worthwhile to mention that results obtained by 
event-by-event analyses are conditional 
on the particular hadronic interaction models used for the 
Monte Carlo simulation. 
The disentanglement of the threefold problem in
determining of the primary mass, 
primary energy and strong interaction features by a combined
analysis can be improved by selecting enriched samples of 
various mass groups. 
Even within one pre-chosen model some hints are expected which 
will enable to understand 
which particular features of the strong interaction models have to 
be improved in order to reproduce the experimental data 
in a consistent way. 

\section{Experimental setup and simulation procedures}

\begin{figure}[ht]
\begin{center}
\vspace*{0.1cm}
\epsfig{figure=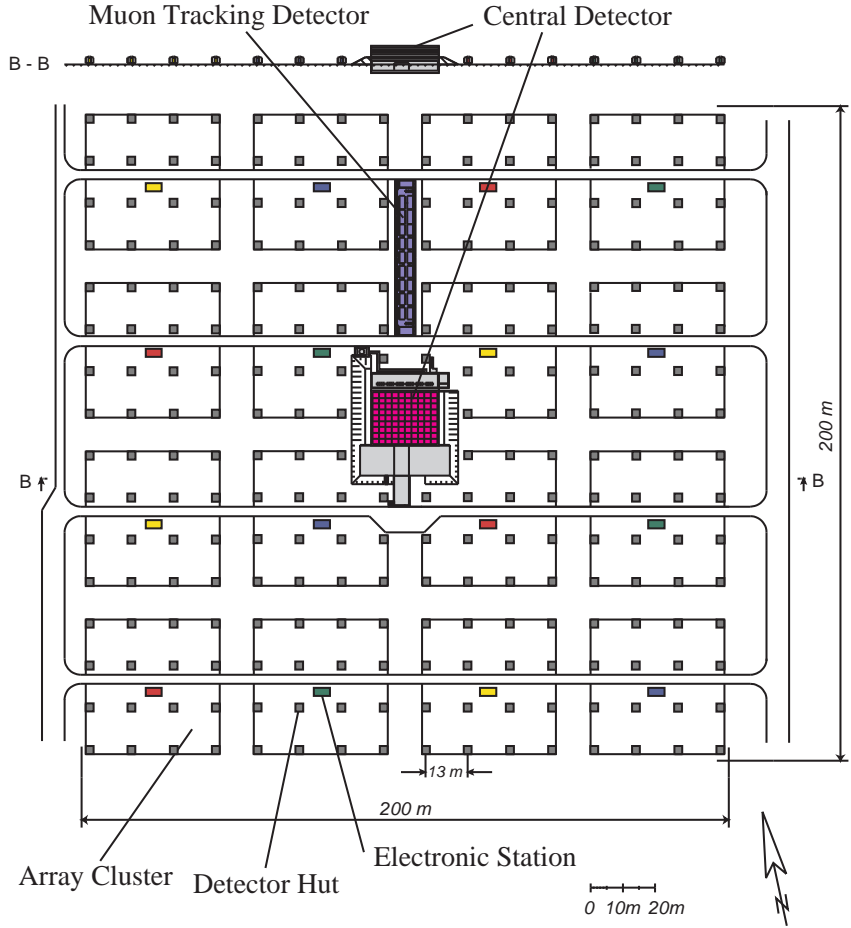,width=0.78\textwidth}
\vspace*{0.1cm}
\end{center}
\caption{\it Layout of the KASCADE experiment.}\label{kascade}
\vspace*{0.3cm}
\end{figure}

The KASCADE experiment, located at the laboratory site of
Forschungszentrum Karlsruhe, Germany at 8$^\circ$E, 49$^\circ$N,
110$\,$m$\,$a.s.l., consists of three main parts - the 
scintillator array, the central detector and the muon tracking 
detector. 
Due to its multi-detector setup, it is able to measure a 
large number of EAS characteristics for each individual event
in the PeV primary energy region.
The schematic view of the KASCADE detector installations 
is shown in Figure~\ref{kascade}.

A scintillator array~\cite{kascade1} measures secondary
electrons, photons and muons 
of extensive air showers in 252 detector stations 
on a grid of 13 m spacing, hence forming an array of 
$200\times200\,$m$^2$. The stations contain unshielded liquid
scintillation counters and below lead and iron absorbers 
also plastic scintillators. With this array the reconstruction of 
the muon and electron size of the EAS is done in an iterative way, 
e.g. by correcting the electron number by use of the measured muon 
content~\cite{lateral}. 

The basic component of the central detector is a 
finely segmented hadron calorimeter~\cite{kascade-cd}. 
A $20\times16\,$m$^2$ iron stack 
arranged in 9 horizontal planes is equipped by liquid ionisation 
chambers forming 44,000 electronic channels. 
The calorimeter measures energy,
angle and point of incidence of individual EAS hadrons.  

Below 30 radiation lengths of absorber the central 
calorimeter contains a layer of 456 scintillation detectors 
\cite{kascade-cd}
acting as trigger for the central detector and measuring the 
arrival time of muons and hadrons. Underneath the 
calorimeter two layers of multiwire proportional chambers (MWPC) 
and one layer of limited streamer tubes (LST)  
reconstruct muon tracks above an energy of 2.4 GeV with 
an angular accuracy of about $1.0^{\circ}\,$~\cite{kascade-mwpc}. 

North of the central detector in a 50 m long tunnel 
muons above the
threshold energy of 0.8 GeV are measured with the help of 
streamer tubes (LST)~\cite{kascade-lst}. On an area of 
about 128$\,$m$^2$ three layers of LST track muons 
with an accuracy of $0.5^{\circ}$.

For the present analysis $\approx\,$700,000 events 
registered by the KASCADE field array are used, corresponding
approximately to one year data taking.
The selection of these showers requires a core distance from the 
center of the array of less than $91\,$m and a
successful reconstruction of the electron size ($N_e$) and 
muon size ($N_\mu^{tr}$). 
The investigated zenith angle range is restricted to 
$15^{\circ}-20^{\circ}$.
In this sample there are around 6000 events where the shower axis
hits the KASCADE central detector, and hence detailed 
hadronic information of the EAS is available for the
analysis.

The simulations for the present analysis use the CORSIKA code 
version 5.62~\cite{corsika} with QGSJET~\cite{ostap1} as
high-energy hadronic interaction model for the EAS development 
in the atmosphere. The options GHEISHA~\cite{GHEISHA} for the interactions at
low energies and EGS4~\cite{EGS} for the electromagnetic cascades are
chosen.

The detailed detector simulation was made on the basis of the 
GEANT~\cite{geant} package, taking into account all shower particles, 
absorber and active materials, energy deposits, 
and arrival times.
More than 20,000 showers are 
generated for each primary nucleus in the primary 
energy range of $5\cdot10^{14} - 3\cdot10^{16}\,$eV. 
The simulations are performed in 10 energy bins with a spectral slope 
of $\gamma=-2.5$ inside and of $\gamma=-1$ from bin to bin.
\begin{table}[!b]
\vspace*{0.5cm}
\caption{\it EAS observables of the 
KASCADE experiment used in the present analysis.}\label{features} 
\begin{center}
\vspace*{0.5cm}
\begin{tabular}{ll}
\hline
$N_{e}$ & Number of electrons in the EAS 
($N_e=2\pi\int_{0}^{\infty}\rho_e(r) r dr$)\\
$N_{\mu}^{tr}$ & Truncated number of muons 
($N_{\mu}^{tr}=2\pi\int_{40{\rm m}}^{200{\rm m}}\rho_\mu(r) r dr$)
($E_\mu > 230\,$MeV)\\
$N_{\mu}^{CD}$ & Number of tracked muons in the central detector 
(MWPC) ($E_\mu > 2.4\,$GeV) \\
$N_{h}$ &  Number of reconstructed hadrons at the calorimeter
($E_{h}>100\,$GeV) \\
$E_{h}^{\rm max}$ & The energy of the most energetic hadron detected
($>100\,$GeV) \\
$E_{\rm tot}$ & Energy sum of the reconstructed hadrons with 
$E_{h}>100\,$GeV \\
\hline
\end{tabular}
\vspace*{0.3cm}
\end{center}
\end{table} 
The simulations cover the angular range of $13^\circ-22^\circ$.
Three different primaries are taken into account: protons, oxygen
nuclei, and iron nuclei.

On this basis different EAS parameters are reconstructed 
whereby simulated and experimental data are 
handled with the same algorithms. 
The EAS core position, arrival direction, electron-muon
densities, electron size and muon content from the array, hadronic 
EAS observables, muon tracks and arrival time distributions observed
with the central detector, 
and many other characteristics are reconstructed.
EAS parameters used in the present analysis 
are compiled in Table~\ref{features}. For a more detailed description
of the reconstruction procedures see e.g. refs.~\cite{roth01,lateral}.

\section{Primary energy and mass determination}\label{energy}

Multivariate methods are used for the classification of the 
measured events in mass groups and for energy estimation. 
These methods take into account the correlations of the 
used observables.
In principle such methods can be applied for any number of 
observables, but the reconstruction quality is restricted 
by the statistical accuracy of the 
reference Monte Carlo sample.    
For the present analysis a multi-layered feed-forward perceptron  
neural network algorithm is used to determine
the mass and energy of individual primary cosmic
rays in the knee region of data registered with KASCADE.
It allows to estimate the primary energy 
and to classify the primary mass into multiple categories 
using similar procedures.
The basics of neural network techniques 
can be found in~\cite{neuro-bishop}.
The general procedures for the application of Bayesian 
and neural network methods at EAS data analysis are given 
in~\cite{ch-neurocomp,roth01}. 

For estimation and classification, the observables of 
the electromagnetic and
muonic components ($N_{e}$,$N_{\mu}^{tr}$) measured by the 
KASCADE field array detectors are used.
The restriction to these two observables is motivated 
by following reasons: \\
$\bullet$ Due to the high statistical accuracy available the 
uncertainties from EAS fluctuations are smaller 
as compared with the hadronic information of EAS. \\
$\bullet$ The information of the KASCADE central detector,
especially the hadronic observables, is intended to be used for
subsequent studies of the interactions of the selected subsamples. \\
$\bullet$ In former studies~\cite{hoerandel,risse,milke} we have found 
that the electromagnetic and muonic component of EAS are described
well by the used high-energy interaction model QGSJET. 

\begin{figure}[ht]
\begin{center}
    \epsfig{file=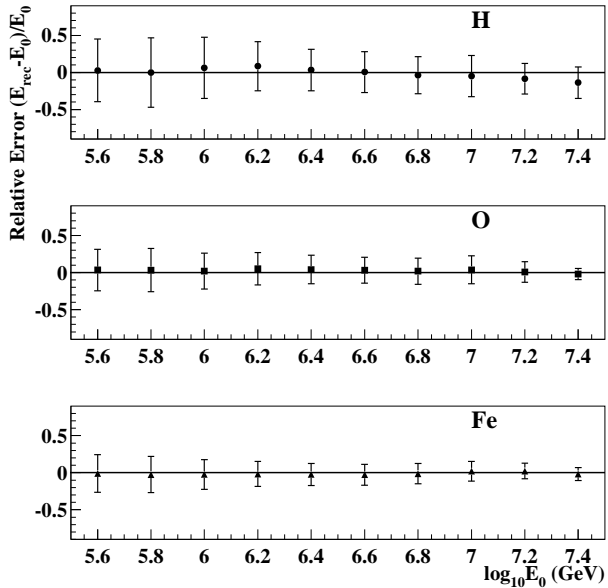,width=0.62\textwidth}
\end{center}
    \vspace{-0.2cm}
    \caption{\it Accuracy of the energy determination for different classes
    of primaries (H, O, Fe) 
    displayed by the relative deviation $(E_{rec}-E_0)/E_0$. The error bars
    indicate the width of the distributions.}
    \label{nenm-bias} 
    \vspace{0.3cm}
\end{figure}

The accuracy of the energy estimation, displayed in 
Figure~\ref{nenm-bias} by the relative deviation of the 
reconstructed energy $E_{rec}$ from the true energy $E_0$ 
results in approximately $25\,$\%, 
with improvements at higher energies and for heavier primaries.
It demonstrates the high reliability nearly free from 
bias eventually arising from the procedure.  
We use a wider energy interval 
for the simulated events than for the experimental ones to 
avoid an over- or underestimation of primary energies at the 
boundaries of the investigated energy region. 
Systematic uncertainties of the energy estimation are 
the composition of the control sample (here three primaries with
equivalent contribution in number of events are used), 
and the high-energy interaction model itself used for the 
generation of the samples.
Of minor influence is the slope of the energy spectrum used at the
Monte Carlo sample if the statistics is large enough over the whole
energy range. 

After estimating the primary energy each EAS event is 
classified as being induced by light, intermediate 
or heavy nuclei. We will refer to these groups as `proton', 
`oxygen', and `iron'. The classification is performed by 
a neural network independently trained from the net used for 
the energy estimation. 
The results of the classification of the trained
neural network are presented in Table~\ref{p2}. 
It shows the probabilities of contamination of events of the different 
classes in each category. The classification matrix is obtained by 
classifying 4000 control events (not used for the training) per class.
A high purity (rather small share of events from alternative classes) 
of proton and iron events is obvious, while the oxygen class has a 
lower purity since it 
\begin{table}[ht]
\vspace*{0.3cm}
\caption{\it Classification probabilities obtained by 
   a neural network classification using a control sample. 
   Used observables are the shower sizes
   $N_{e}$ and $N_{\mu}^{tr}$. $W_{\rm i\longleftarrow j}$
   denotes the abundance of events of type ${\rm j}$ of the 
   sample classified as ${\rm i}$.}\label{p2} 
\vspace*{0.3cm}
 \begin{center}  
  \begin{tabular}{|l|ccc|}
      \hline
      $W_{\rm i\longleftarrow j}$ & j=H $[\%]$ & j=O  $[\%]$ & j=Fe $[\%]$ \\
      \hline
      i=H & 80 &  18 &  2\\
      i=O & 19 &  58 &  23\\
      i=Fe & 2  & 23 &  75\\
      \hline
    \end{tabular}
\vspace*{0.3cm}
 \end{center}
\end{table}
contains a significant contamination from both, protons and irons.
The classification depends slightly on primary energy with 
improved accuracy at higher energies by $\approx 10$\% due to 
decreasing fluctuations of the observables. 
The restriction to three mass groups
leads to systematic distortions if intermediate primary mass groups
are present in control or measured samples. 
For example, helium nuclei would be classified mainly as protons 
and a part of them (more than protons) would be attributed to the 
medium and heavy classes (see ref.~\cite{roth01}).
 
\begin{figure}[!b]
  \centering
    \vspace{.3cm}
  \begin{minipage}[t]{0.47\textwidth}
    \begin{center}
    \epsfig{file=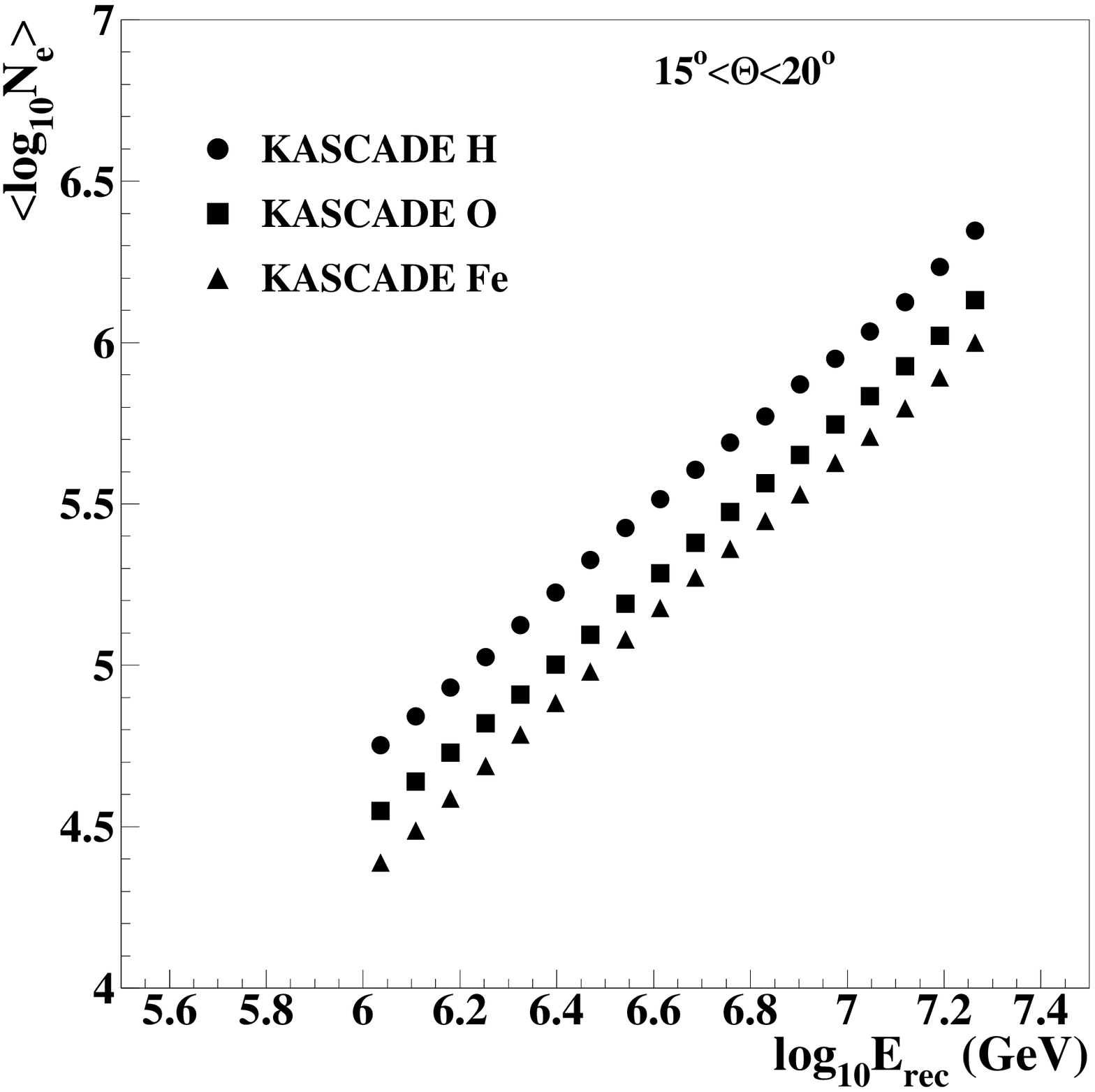,width=0.99\textwidth}
    \end{center}
    \vspace{-.2cm}
    \caption{\it Mean shower size $N_e$ versus the reconstructed 
       primary energy $E_{rec}$ for the
       measured KASCADE data set, classified in proton,
       oxygen, and iron samples.}\label{ne-exp-pof}
  \end{minipage}
  \hspace{0.04\textwidth}
  \begin{minipage}[t]{0.47\textwidth}
    \begin{center}
    \epsfig{file=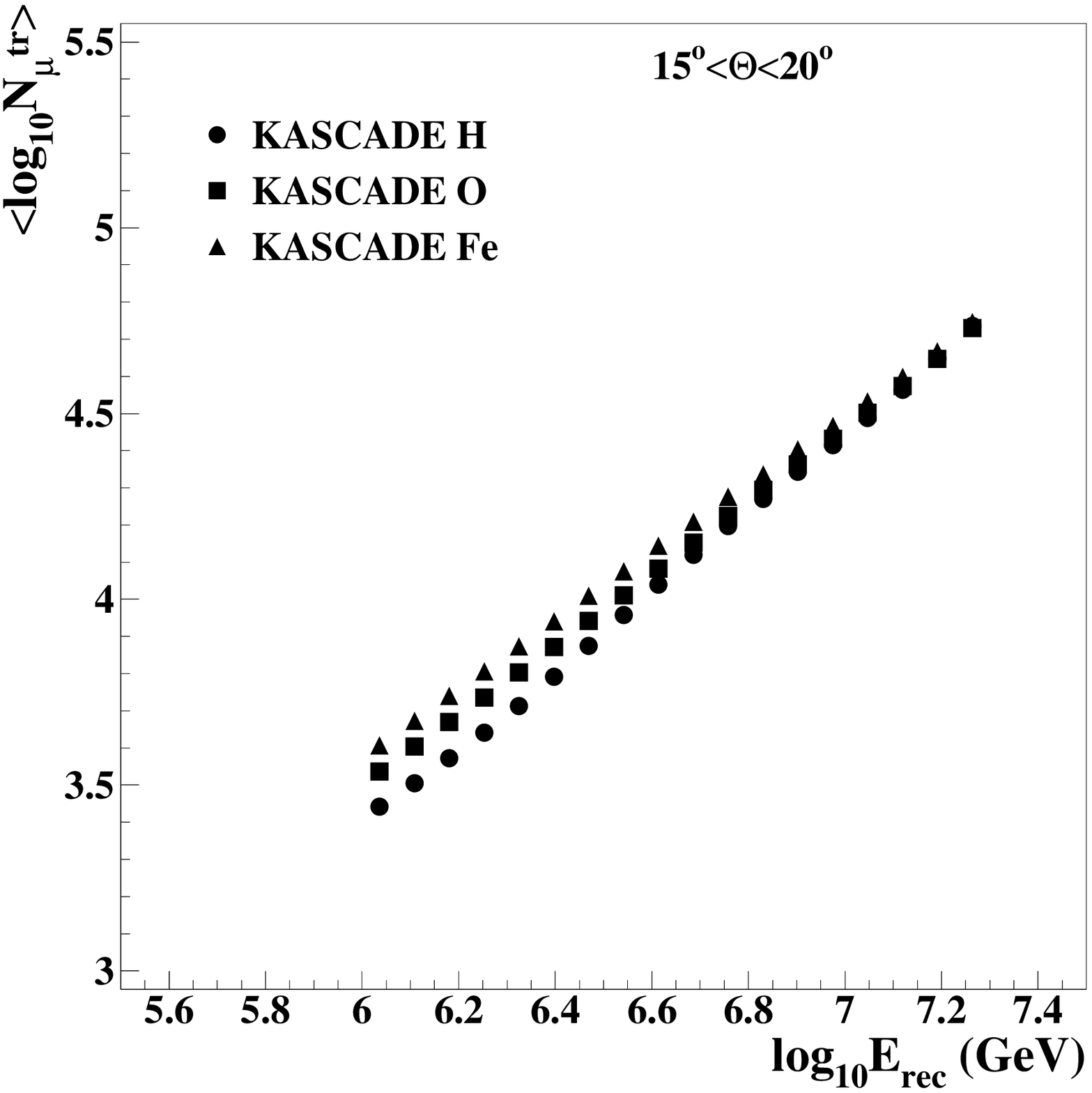,width=0.99\textwidth}
    \end{center}
    \vspace{-.2cm}
    \caption{\it Mean muon size $N_{\mu}^{tr}$ versus the 
       reconstructed primary energy $E_{rec}$ for the
       measured KASCADE data set, classified in proton,
       oxygen, and iron samples.}\label{lnmu-exp-pof} 
  \end{minipage}
\vspace{0.3cm}
\end{figure}

After applying the trained neural networks to measurements we 
combine the energy and mass information of the analyzed KASCADE
data sample. Figures~\ref{ne-exp-pof} and~\ref{lnmu-exp-pof} 
display the ($N_e$,$E_{rec}$)- and ($N_{\mu}^{tr}$,$E_{rec}$)-dependence 
for the three selected samples of primaries.
The energy resolution is expected to be $\sim 25\,$\% as demonstrated 
by Monte Carlo simulations (Fig.~\ref{nenm-bias}), and the mass 
discrimination power is $\sim 70\,$\% as illustrated in Table~\ref{p2}.
The discrimination power is defined as the geometric mean of
the probabilities $W_{\rm i\longleftarrow i}$. 
It is obvious that the mean $N_{e}$ values 
are rather close for the intermediate and heavy groups of nuclei, 
which explains the comparatively strong mixture between these 
two classes 
($W_{\rm Fe \rightarrow O} = W_{\rm O \rightarrow Fe} = 23\%$). 
The mean numbers of muons are approximately the same for all 
primaries in all energy bins. One recognizes small differences 
from that
in the lowest energy bins, but the overall independence of 
$N_{\mu}^{tr}$ from the primary mass is obvious, i.e.
$N_{\mu}^{tr}$ dominates (at KASCADE observation 
level) the energy estimation.
The slight deviations at high energies from a pure power law in case
of protons are probably due to a small underestimation 
at highest energies ($E_{0}>2\cdot10^{16} eV$) 
(see Figure~\ref{nenm-bias}).

\section{The purification procedure}\label{pp}

\begin{figure}[!h]
  \centering
    \vspace{0.5cm}
  \begin{minipage}[t]{0.47\textwidth}
    \epsfig{file=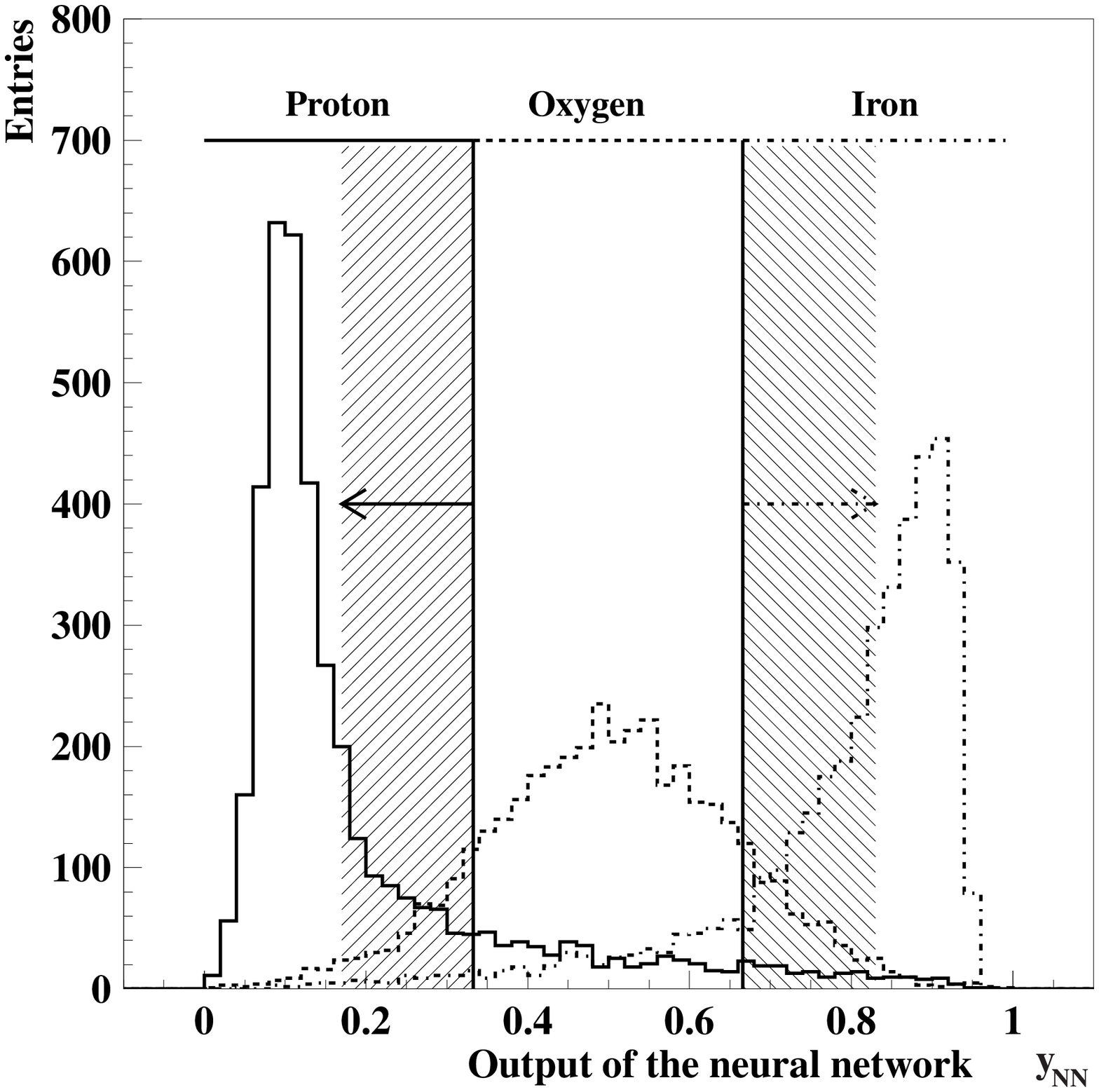,width=1.02\textwidth}
    \vspace{-0.5cm}
    \caption{\it The distribution of the neural network output for the 
    simulated control sample. Purification can be performed by 
    shifting the boundaries of the subintervals.}\label{nnout}
  \end{minipage}
  \hspace{0.04\textwidth}
  \begin{minipage}[t]{0.47\textwidth}
    \epsfig{file=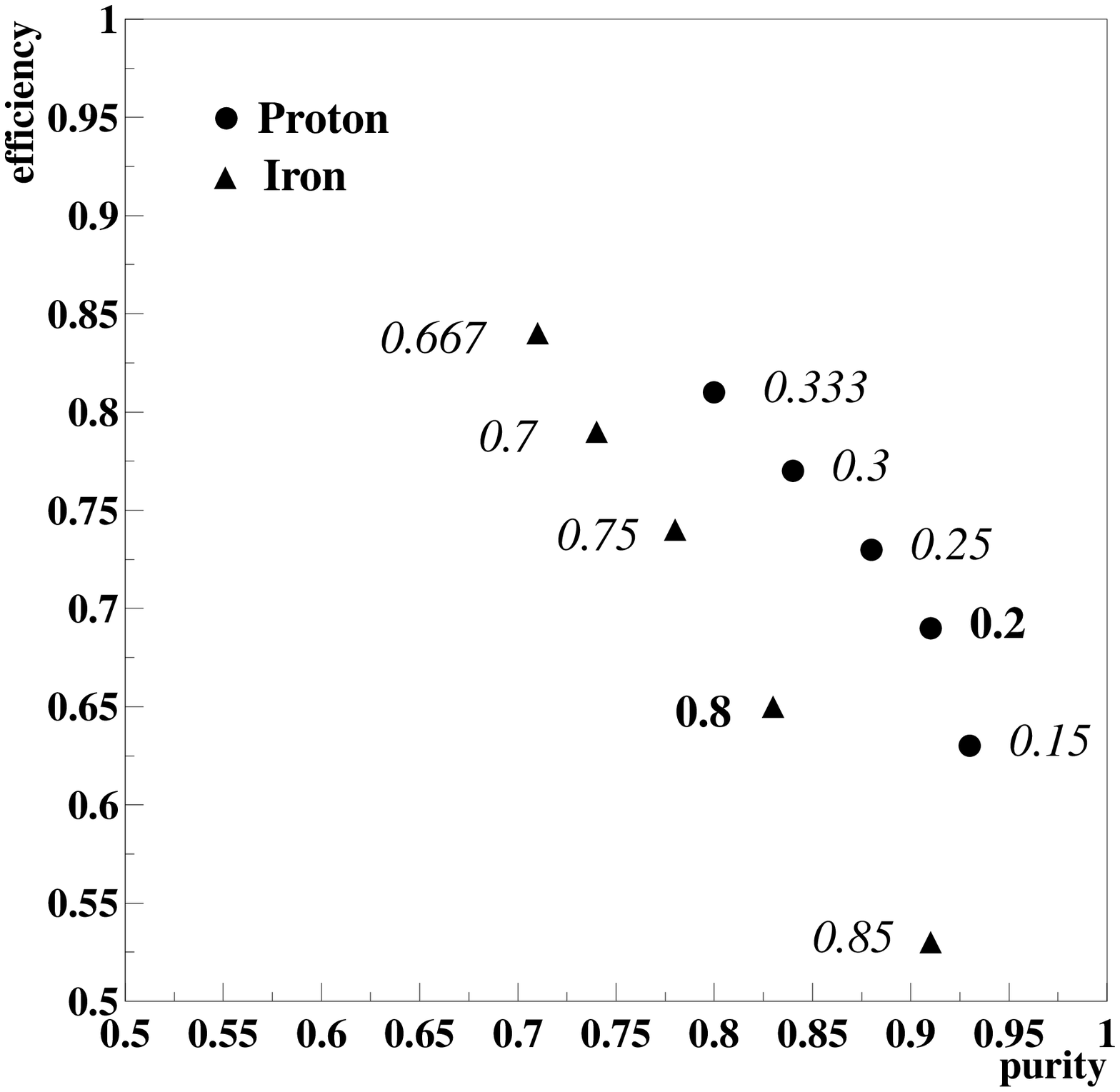,width=1.02\textwidth}
    \vspace{-0.5cm}
    \caption{\it Efficiency vs. purity for proton and 
    iron events by variation of the subinterval boundaries
    as indicated by the numbers. The values are
    obtained by classifying a Monte Carlo 
    control sample with equal total numbers of 
    primaries for each mass.}\label{pure}
  \end{minipage}
\vspace{0.3cm}
\end{figure}

The neural network analyses perform a nonlinear mapping of 
multidimensional characteristics of the EAS to 
the real number interval $[0,1]$ (Fig.~\ref{nnout}).
Particular class assignments for the three way classification 
are the subintervals $[0.,0.33)$, $[0.33,0.66]$ and $(0.66,1.]$ for
the light, medium, and heavy nuclei, respectively.
We characterize the quality of the classification procedures by the 
`purity' and `efficiency' variables. 
The purity of a sample is defined as the fraction
of true classified events in an actual number of events assigned to 
a given class.
The classification efficiency 
is defined as the fraction of true classified events to the
initial number of events of a given class.
The actual classification procedure results in a purity
of 80\% for the proton class, and of 70\% for the iron class
assuming equal total numbers of primaries in each of the three 
classes. 
The neural information technique~\cite{ani} allows now to 
reduce the contamination of misclassified events in each class 
of nuclei. 
Of course, the efficiency 
of the classification is reduced at the same time.
The optimum of purity and efficiency to be chosen depends on the 
given problem. Investigations of the behavior of definite primaries
requires a higher purity at may be efficiency,
whereas estimation of chemical composition needs high
efficiency.
It should be remarked, that for obtaining results on chemical composition
of a measured sample the numbers in each class have to be corrected
with the misclassification matrix. Changing the boundaries of the
class assignments will always result in the same composition 
after the correction, if no systematic effects are introduced. 
In this analysis 
the possibility of the
selection of maximally pure samples of cosmic ray mass groups 
will now be investigated,
with respect to the question of the cost we have to pay 
(in terms of efficiency loss) to get light and heavy nuclei 
induced showers with higher purity. 

When the neural network (NN) 
is satisfactorily trained, the NN output distributions
for the different classes are overlapping at the subinterval
boundaries. Therefore, by shrinking the subintervals, one can 
remove a large proportion of misclassified events. But,
simultaneously one looses parts of the correctly classified events.
Figure~\ref{nnout} illustrates this procedure of purification.

Figure~\ref{pure} plots purity versus efficiency for 
two classes. For equal total number of simulated events 
the purity of proton and 
iron nuclei can reach more than $90\,$\% 
while the efficiency is still remaining above $50\,$\%. 
The purity and efficiencies are obtained by classifying 
4000 simulated control events per class which are 
not used for the training of the neural network. 
For a given purity value the efficiency of proton 
events is always slightly larger than the efficiency 
for the iron induced sample, due to the narrower NN output 
distribution of the protons (Fig.~\ref{nnout}), e.g. the
separation of oxygen-proton is better than oxygen-iron. 
The separation of protons from the other classes is good due to
the combination of proton and helium nuclei in one class.
%
\begin{figure}[ht]
\begin{center}
\vspace{.2cm}
    \epsfig{file=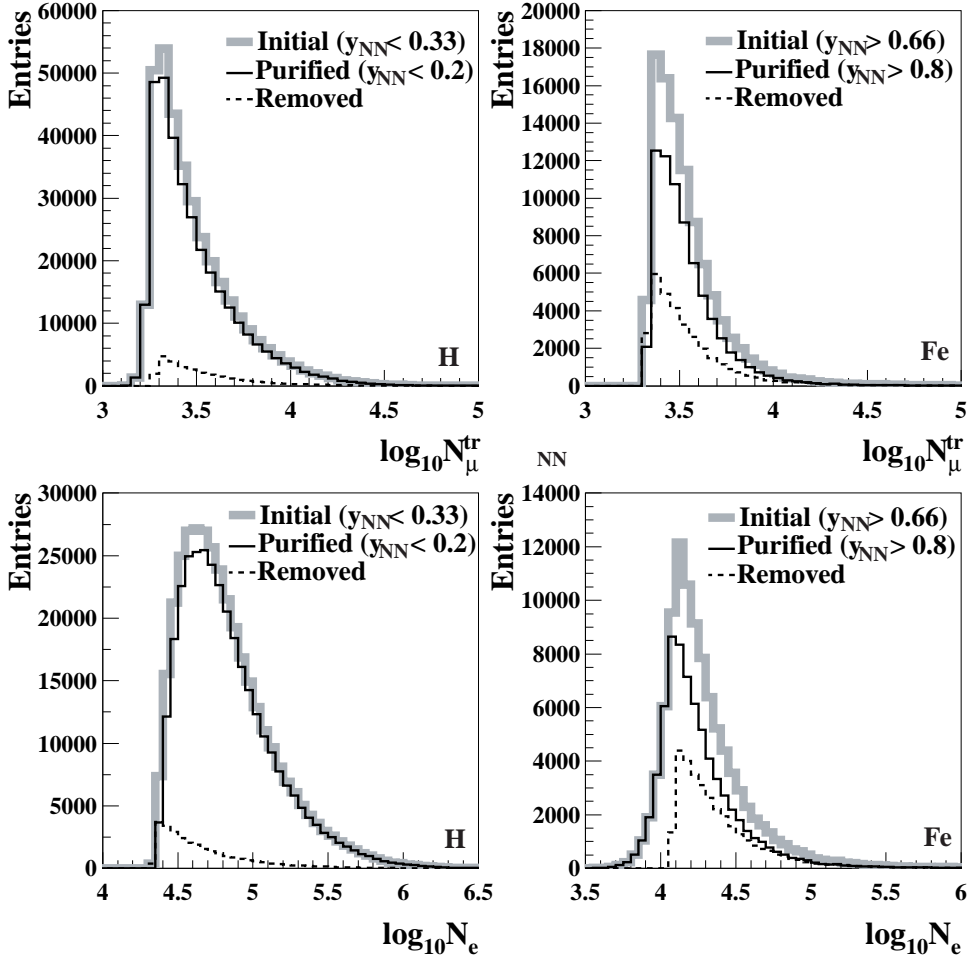,width=0.9\textwidth}
  \end{center}
\vspace{0.2cm}
    \caption{\it Distributions of $N_e$ and $N_{\mu}^{tr}$
    for the initial, the purified, and removed proton and
    iron samples as measured with KASCADE. Details 
    see text.}\label{cuts-ne-nm} 
\vspace{0.5cm}
\end{figure}

For the preparation of the enriched samples to 
investigate the hadronic interactions the
\begin{figure}[ht]
\begin{center}
\vspace{.2cm}
    \epsfig{file=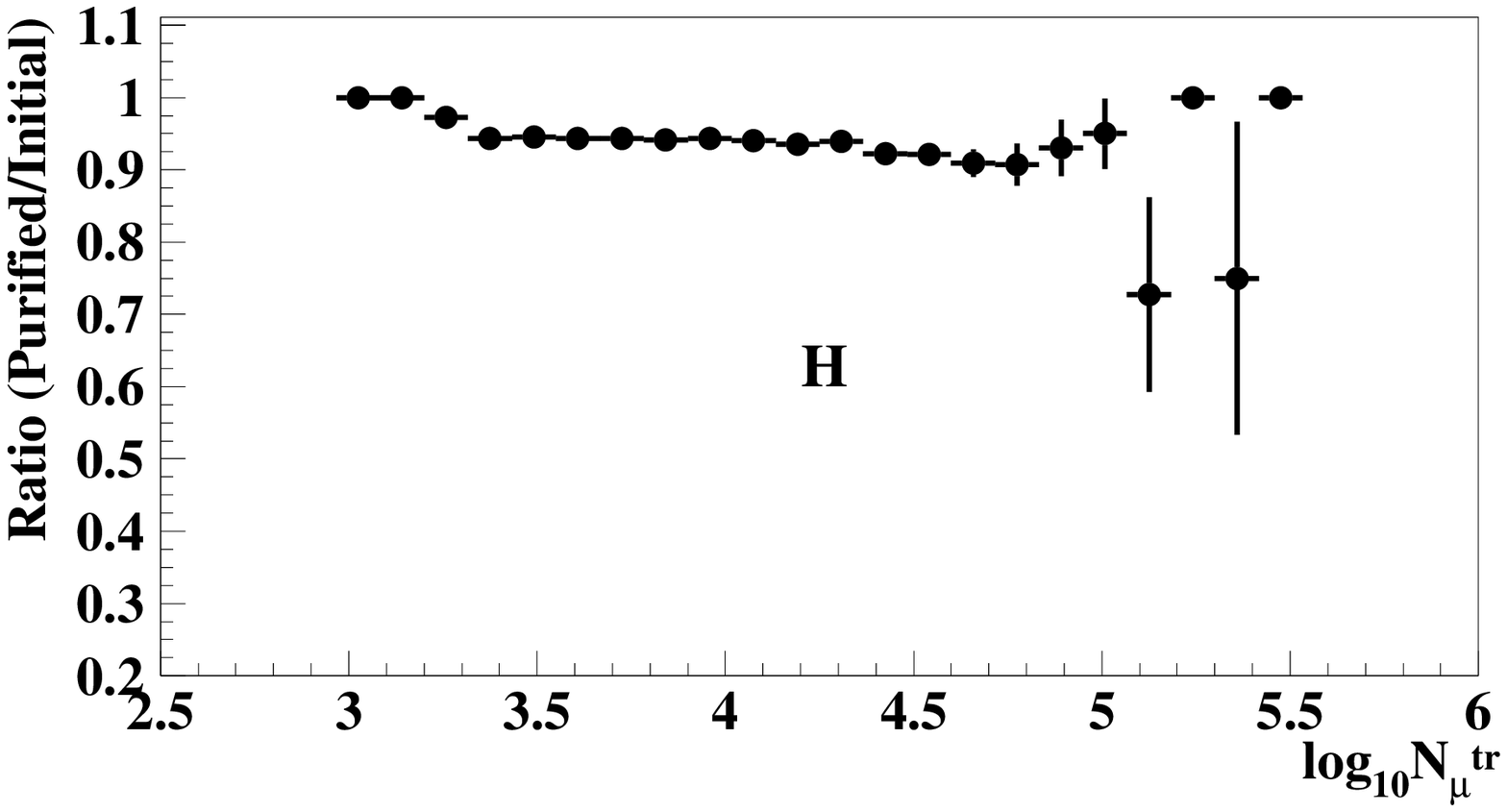,width=0.6\textwidth}
\vspace{-0.1cm}
  \end{center}
    \caption{\it The ratio of the purified to initial distribution 
    for $N_{\mu}^{tr}$ of the proton sample as measured with 
    KASCADE.}\label{pirat} 
\vspace{0.5cm}
\end{figure}
purification procedure has to be scrutinized in order to 
get the optimal purification not to distorte the initial 
parameter distributions. 
Hence, first of all we investigate how the shrinking 
of the interval for the different classes at the NN output  
affects the corresponding one- and two-dimensional distributions 
of the selected events, i.e. the bias introduced by the 
purification.
For this we compare the distributions of observables for the 
measured EAS samples of
events selected by the purification with the removed ones.
Figure~\ref{cuts-ne-nm} shows the distributions of EAS parameters 
of proton and iron classes for two different cuts in the 
NN output intervals applied to the KASCADE data sample. The upper two 
plots show the purified and removed distributions 
compared with the initial ones in case of the muon size, and
the lower two plots in case of the electron number.
For the shown purification the boundaries are
shifted from y$_{\rm NN}=0.33$ to y$_{\rm NN}=0.2$ and from 
y$_{\rm NN}=0.66$ to y$_{\rm NN}=0.8$, respectively. 
\begin{table}[ht]
\vspace*{0.2cm}
  \begin{center}
    \caption{\it The mean values $\mu$ (and its variances) of parameter 
    distributions (Fig.~\ref{cuts-ne-nm}) of the 
    initial, purified, and removed proton and iron samples of the
    KASCADE experimental data set.}\label{mean-val}
\vspace*{0.5cm}
    \begin{tabular}{|l||c|c|c|c|}
      \hline
      \hline
 & $\mu_{N_{e}}^{\rm H}$$(\sigma)$ 
 & $\mu_{N_{e}}^{\rm Fe}$$(\sigma)$
 & $\mu_{N_{\mu}^{tr}}^{\rm H}$$(\sigma)$ 
 & $\mu_{N_{\mu}^{tr}}^{\rm Fe}$$(\sigma)$ \\
      \hline
       Initial     & 4.82(0.316) & 4.30(0.311) & 3.49(0.228) & 3.54(0.210)\\
       Purified    & 4.83(0.313) & 4.26(0.300) & 3.49(0.227) & 3.54(0.202)\\
       Removed     & 4.67(0.309) & 4.40(0.316) & 3.53(0.237) & 3.55(0.229)\\
      \hline
    \end{tabular}
  \end{center}
\vspace*{0.3cm}
\end{table}
The events are removed over nearly the whole 
range of the distributions proving the small dependence of the
classification on primary energy.
A more detailed inspection of Figure~\ref{cuts-ne-nm} shows that only few 
events with smallest shower size $N_{e}$ are removed from the iron sample and 
only few events with largest size from the proton
events distribution. The opposite situation is observed for the 
$N_{\mu}^{tr}$ distribution.
Figure~\ref{pirat} illustrates this feature by displaying as 
example the ratio of the purified to initial sample of the 
$N_{\mu}^{tr}$ distributions for the proton sample. 
The shape (Fig.~\ref{cuts-ne-nm}) and the mean values
$\mu_{N_i}^{j}$ with $i\in{e,\mu}$ and $j\in{\rm H,Fe}$
(Table~\ref{mean-val}) of the distributions are only slightly 
changed by the purification procedure. 
For further investigations using the purified samples 
it is important that they still represent the total 
range of the distributions. 

In order to scrutinize possible systematic 
distortions of distributions 
of parameters not used in the classification procedure,
one-dimensional statistical tests~\cite{fukunaga}
have been performed comparing the initial and the purified samples. 
Table~\ref{od-tests} displays the results of 
three different tests for 
hadronic observables measured with the 
KASCADE central detector.
\begin{table}[!b]
\vspace*{0.2cm}
  \begin{center}
    \caption{\it The probability values of different tests ($t$- Student, 
    $KS$- Kolmogorov-Smirnov, $MW$- Mann-Whitney) comparing the
    initial and purified proton and iron samples of various hadronic
    shower observables.}\label{od-tests}
\vspace*{0.5cm}
    \begin{tabular}{|l|c|c|c||c|c|c|}
    \hline
 &  \multicolumn{3}{c||}{proton distributions} &
    \multicolumn{3}{c|}{iron distributions} \\
    \cline{2-7}
       & $t$ & $KS$ & $MW$ & $t$ & $KS$ & $MW$ \\
    \hline
    $N_{h}$        & 0.36 & 0.99 & 0.49 & 0.35 & 0.99 & 0.28 \\
    $E_{h}$        & 0.36 & 0.77 & 0.40 & 0.38 & 0.72 & 0.23 \\
    $E_{h}^{max}$  & 0.30 & 0.81 & 0.29 & 0.40 & 0.84 & 0.26 \\
      \hline
    \end{tabular}
  \end{center}
\vspace*{0.3cm}
\end{table}
We perform different tests because
they compare the distributions using varying 
statistical criteria.
The presented values are the probabilities of accepting  
the null hypothesis, which is that these two samples are from one and 
the same population. 
In cases of small probabilities the null hypothesis
is rejected, i.e. there exists a big difference 
between the two samples.
It was found that all probabilities from Table~\ref{od-tests} 
are above the critical values of these 
tests to reject the null hypothesis~\cite{fukunaga}. 
The initial and purified proton and iron samples 
belong statistically to the same population demonstrating that the 
purification does not introduce large systematic distortions.
It should be noted that these tests have been done for hadronic
observables which are not used for energy estimation and mass 
classification. 

Generally, the one-dimensional distributions are not strongly  
affected by more detailed cuts in the NN output distribution. 
Therefore, a cut applied to the NN output distribution is not 
linearly transfered to the distributions of the input parameters. 
\begin{figure}[!b]
\vspace{0.3cm}
\begin{center}
    \epsfig{file=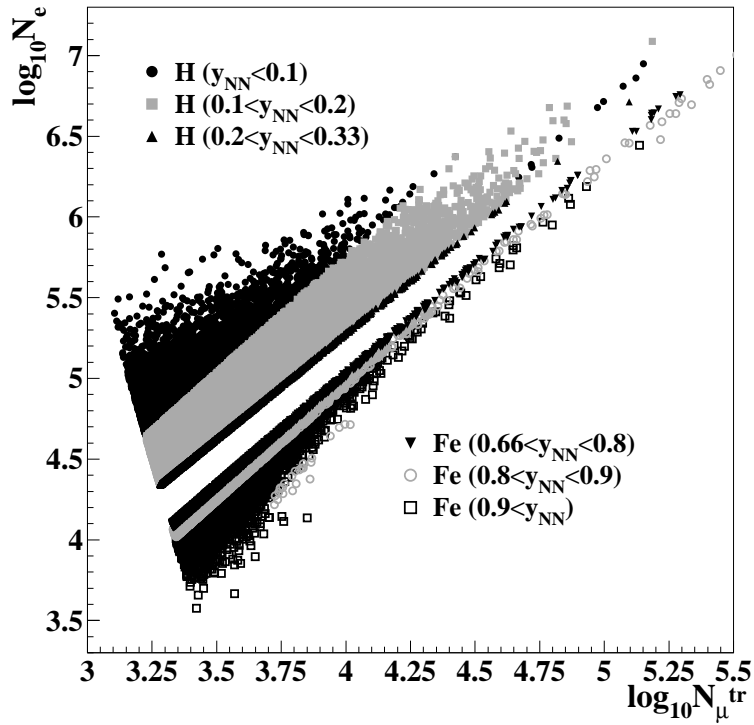,width=0.75\textwidth}
\vspace{-.2cm}
    \caption{\label{ne-lnmu} \it Two-dimensional distribution 
    of the NN input
    parameters $N_e$ and $N_{\mu}^{tr}$ with different cuts applied on
    the NN output parameter. The cut values are indicated.} 
\end{center}
\vspace{0.3cm}
\end{figure}
This feature illustrates that the neural network analysis performs a 
nonlinear mapping of the multidimensional input vector to 
an one-dimensional output value, with
the observation that the fraction of removed
events at the $N_{e}$- and $N_{\mu}^{tr}$-distributions
is nearly independent of the shower sizes.
This is demonstrated in Figure~\ref{ne-lnmu}, 
which displays the ($N_{e}$,$N_{\mu}^{tr}$)-parameters as a 
two-dimensional distribution. It is obvious that we remove both 
proton and iron events from the boundary region, 
where the misclassification probability is high.
But by changing the cut values in the NN output different
regions of the ($N_{e}$,$N_{\mu}^{tr}$) parameter space
are involved. 
The nonlinearity of the mapping is of importance as linear cuts 
would not do justice to the intrinsic shower fluctuations in 
$N_{e}$ and $N_{\mu}^{tr}$ which depend on primary energy and mass.

We conclude that only small systematic uncertainties 
are introduced to the parameter distributions when 
`purifying' the proton and iron samples. But there are still 
other parameters which may be systematically affected by the 
cuts.  
\begin{figure}[ht]
\vspace{0.3cm}
\begin{center}
    \epsfig{file=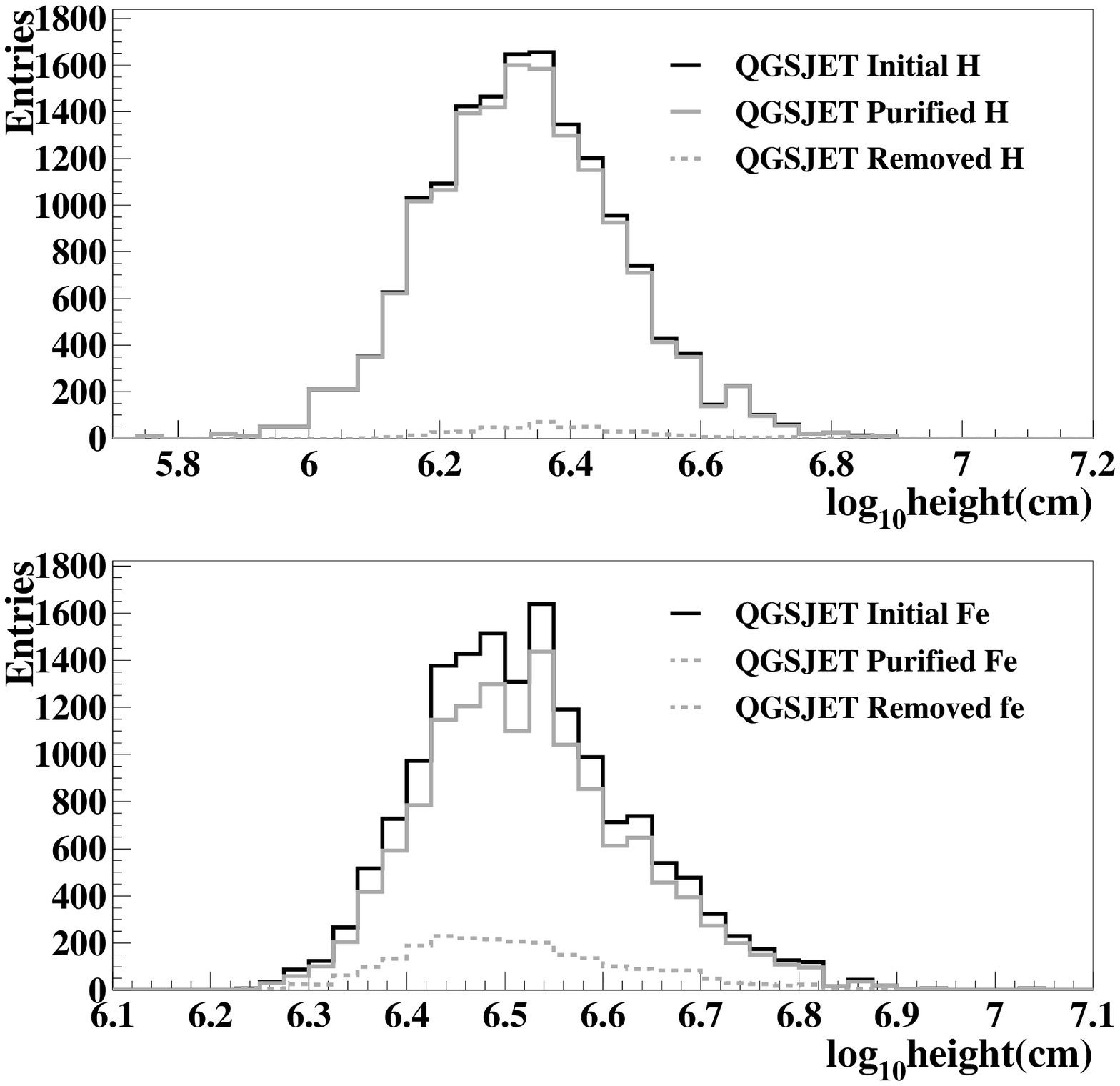,width=7.cm}
    \epsfig{file=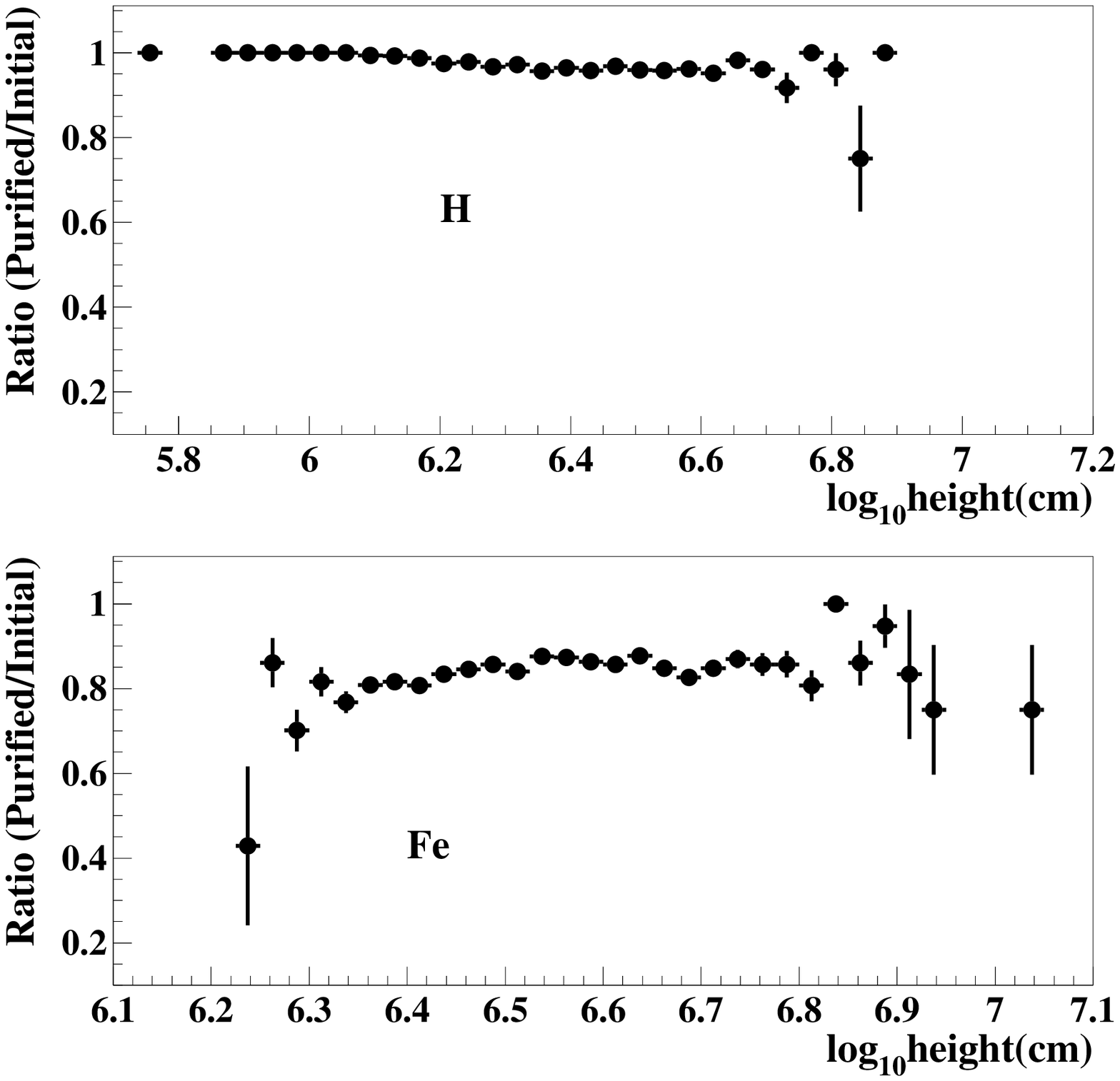,width=7.cm}
\end{center}
    \caption{\it Same as Fig.~\ref{cuts-ne-nm}, but for the
    height of the first interaction and in case of
    Monte Carlo simulations. In the right part the ratio
    of the purified to initial samples 
    are displayed.}\label{cuts-mcz} 
\end{figure}
One of these parameters is the height of the first interaction 
of the primary nucleus. 
Unfortunately this parameter is not accessible by our experimental
data. Therefore the distributions of this parameter for initial and 
purified proton and iron samples have been only investigated 
for MC data.
As the height of the first interaction influences directly the
muon and electron shower sizes at observation level,
we expect that the purification affects mainly the 
boundary region (low heights for iron and large heights for 
proton) of the height distribution.
This is confirmed as displayed in Figure~\ref{cuts-mcz}. Additionally 
Figure~\ref{cuts-mcz} shows, 
that an eventual systematic bias introduced for this parameter 
remains again small.

\section{Studies with mass enriched EAS samples}

An estimation of the primary energy with $\sim 25\,$\% 
relative error in addition to 
an accurate classification of the primary particles into 3 mass 
categories was performed.
The purification technique was applied for preparing  
the enriched samples 
using only the KASCADE array information ($N_e, N^{tr}_{\mu}$)
(see also~\cite{ch-ani99-2}).
The procedure facilitates production of 
enriched samples of proton and iron
induced events with the additional knowledge of the primary energy
on an event-by-event basis.
Some characteristic features of these purified
`light' and `heavy' samples
will now be discussed, especially the behavior of hadronic
observables with primary energy. 
The event selection and reconstruction 
procedures are done for simulated and experimental data samples
in a corresponding way. 
\begin{figure}[h]
    \vspace{0.3cm}
  \centering
  \begin{minipage}[t]{0.47\textwidth}
    \epsfig{file=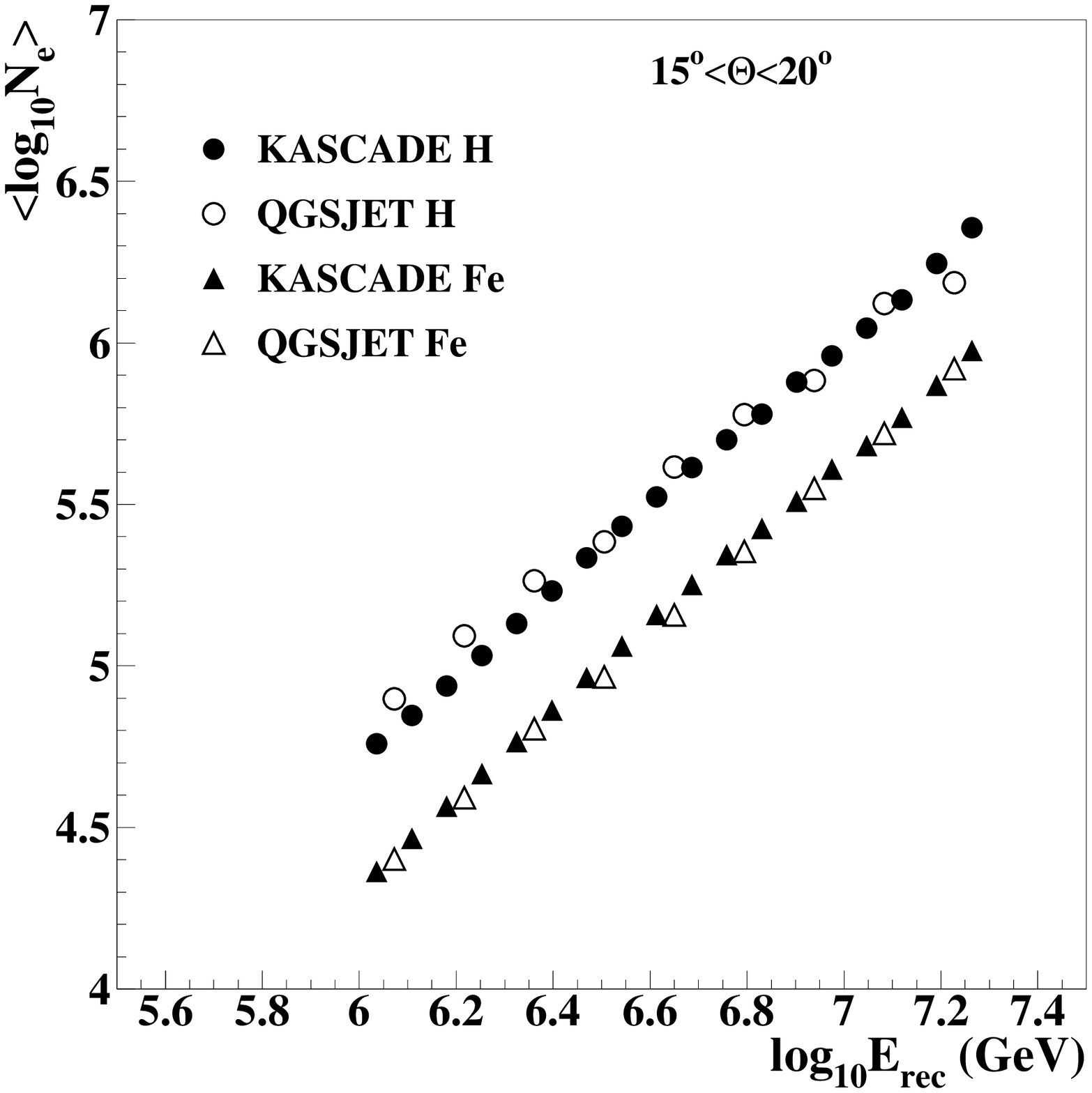,width=0.95\textwidth}
    \vspace{-0.2cm}
    \caption{\it Shower size $N_e$ versus the reconstructed energy 
    $E_{rec}$  for simulated and
     experimental proton and iron events. The primary
      energy is estimated by neural
      regression method.}\label{ne-pf} 
  \end{minipage}
  \hspace{0.04\textwidth}
  \begin{minipage}[t]{0.47\textwidth}
      \epsfig{file=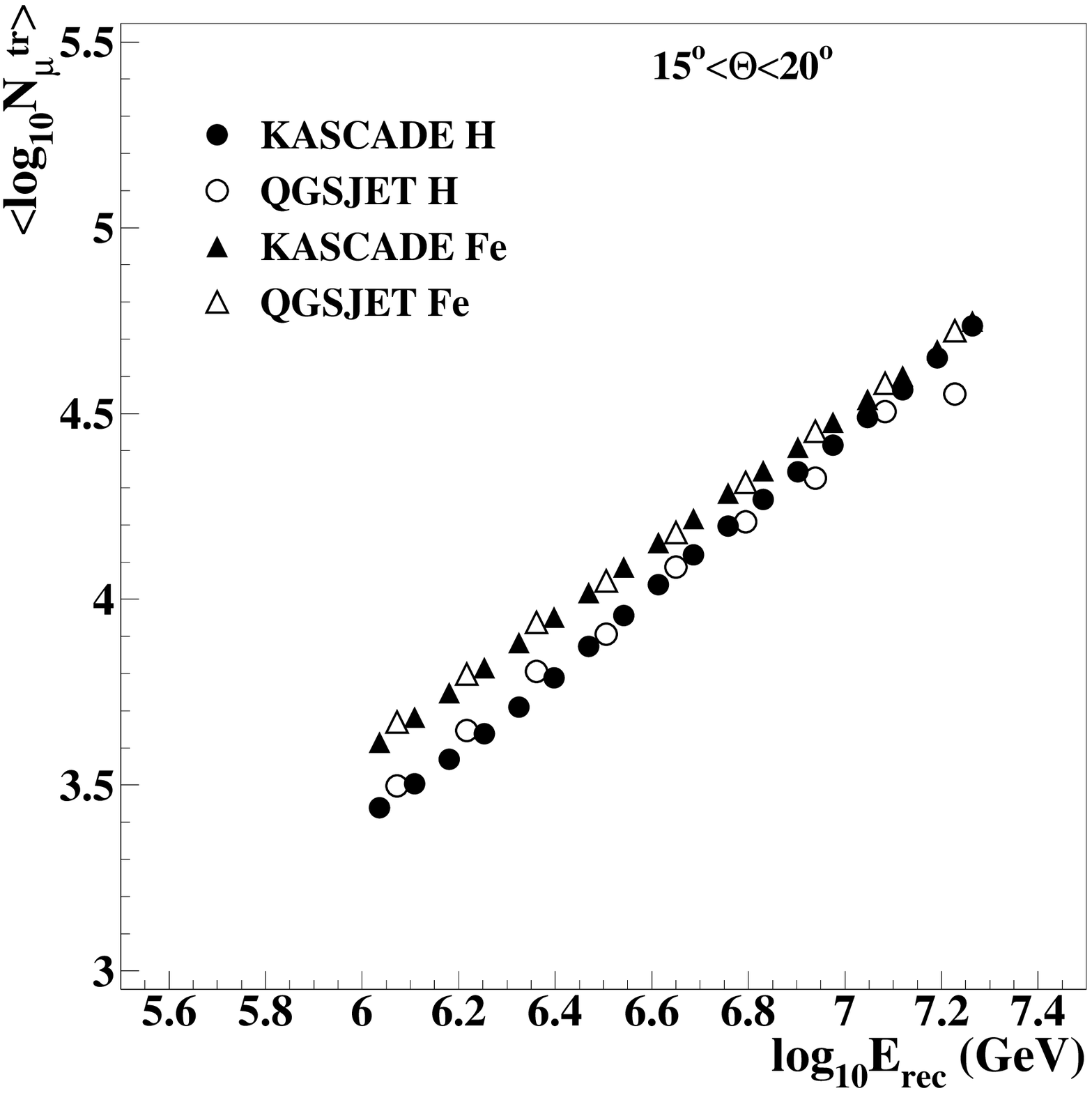,width=0.95\textwidth}
    \vspace{-0.2cm}
    \caption{\it Muon size $N_{\mu}^{tr}$ versus the reconstructed
    energy $E_{rec}$ for simulated and 
       experimental proton and iron events. The primary energy 
      is estimated by neural regression method.}\label{lnmu-pf}
  \end{minipage}
\vspace{0.5cm}
\end{figure}
First, the parameters $N_e$ and $N^{tr}_{\mu}$ of the 
experimental events are compared with those from MC simulations  
(Figures~\ref{ne-pf} and \ref{lnmu-pf}).
A good agreement is displayed, 
demonstrating the high methodical accuracy, and  
furthermore, that the electromagnetic and muonic 
components of EAS are well described by the MC model used. 
As the method takes also the
correlation of the two observables into account, the agreement 
in both observables suggests the validity of the Monte Carlo 
model for these gross shower parameters. 

The purification of 
the  samples makes it possible to study
high-energy muons and hadronic observables detected 
by the KASCADE central detector separately for showers
induced by different primary mass. These parameters are not 
used for energy estimation and mass clasiification and for
producing the enriched samples.
In this context a test can be performed
on the balance of energy and particle number distributions 
of the different shower components in the Monte Carlo model. 
For illustration purposes of the presented techniques we consider the
energy sum $E_{\rm tot}$ of the hadrons with $E_h>100\,$GeV  
reconstructed on shower-to-shower basis
from calorimeter data of the central detector system.
Only EAS with the core inside the area of the 
central detector have been used. Additionally an electron number of 
more than 10,000 and at least two reconstructed hadrons 
($E_h>100\,$GeV) for a single event were required. 
Hence, the number of selected events is noticeably reduced.

\begin{figure}[ht]
\vspace{0.5cm}
\begin{minipage}[t]{0.47\textwidth}
	\epsfig{file=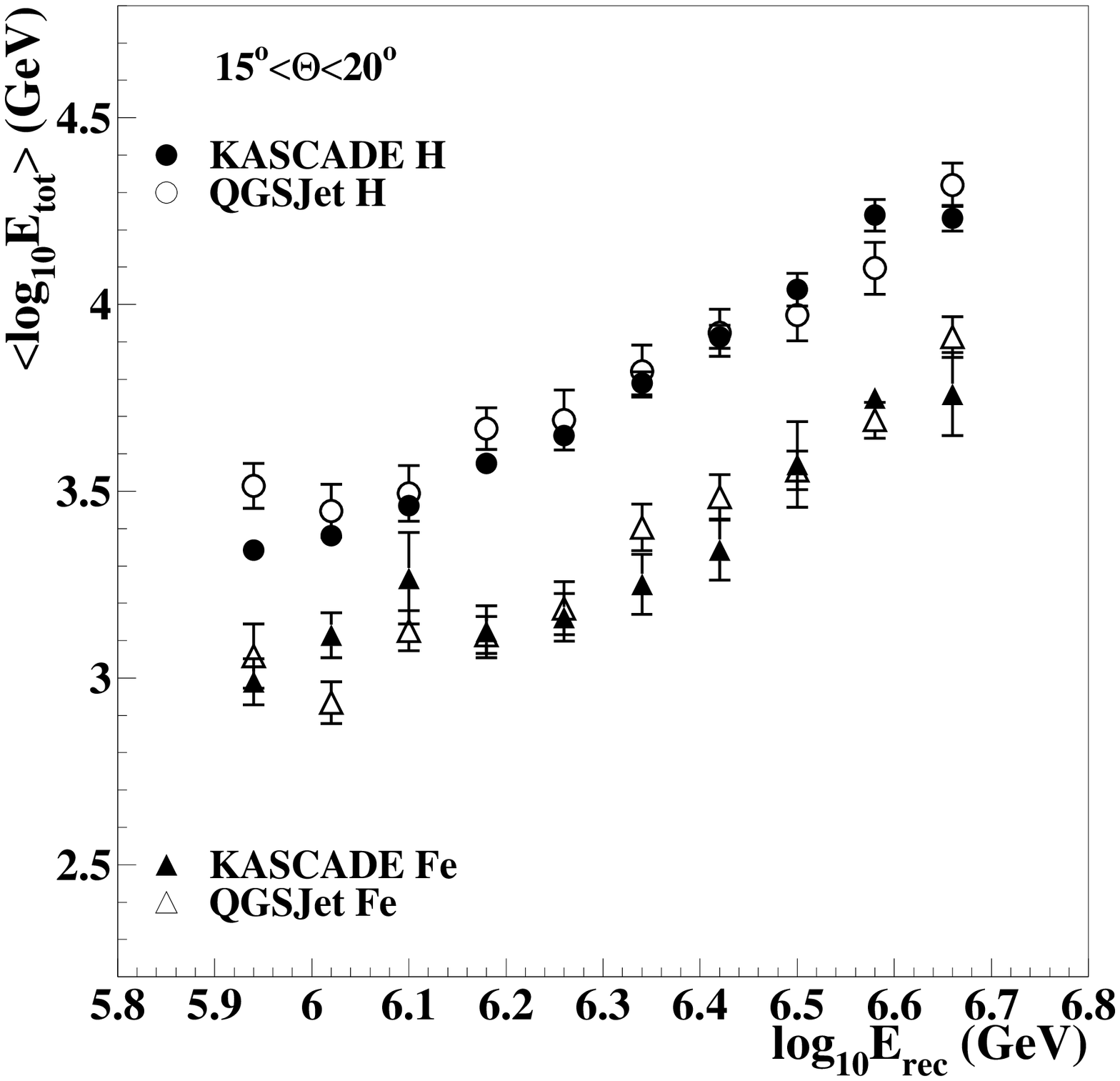,width=0.99\textwidth}
\vspace{-0.9cm}
\caption{\label{eh} \it{Dependence of the energy sum of
reconstructed hadrons $E_h>100\,$GeV on the reconstructed 
primary energy for simulated and experimental proton and 
iron classes.}}
\end{minipage}
\hfill
\begin{minipage}[t]{0.47\textwidth}
\epsfig{file=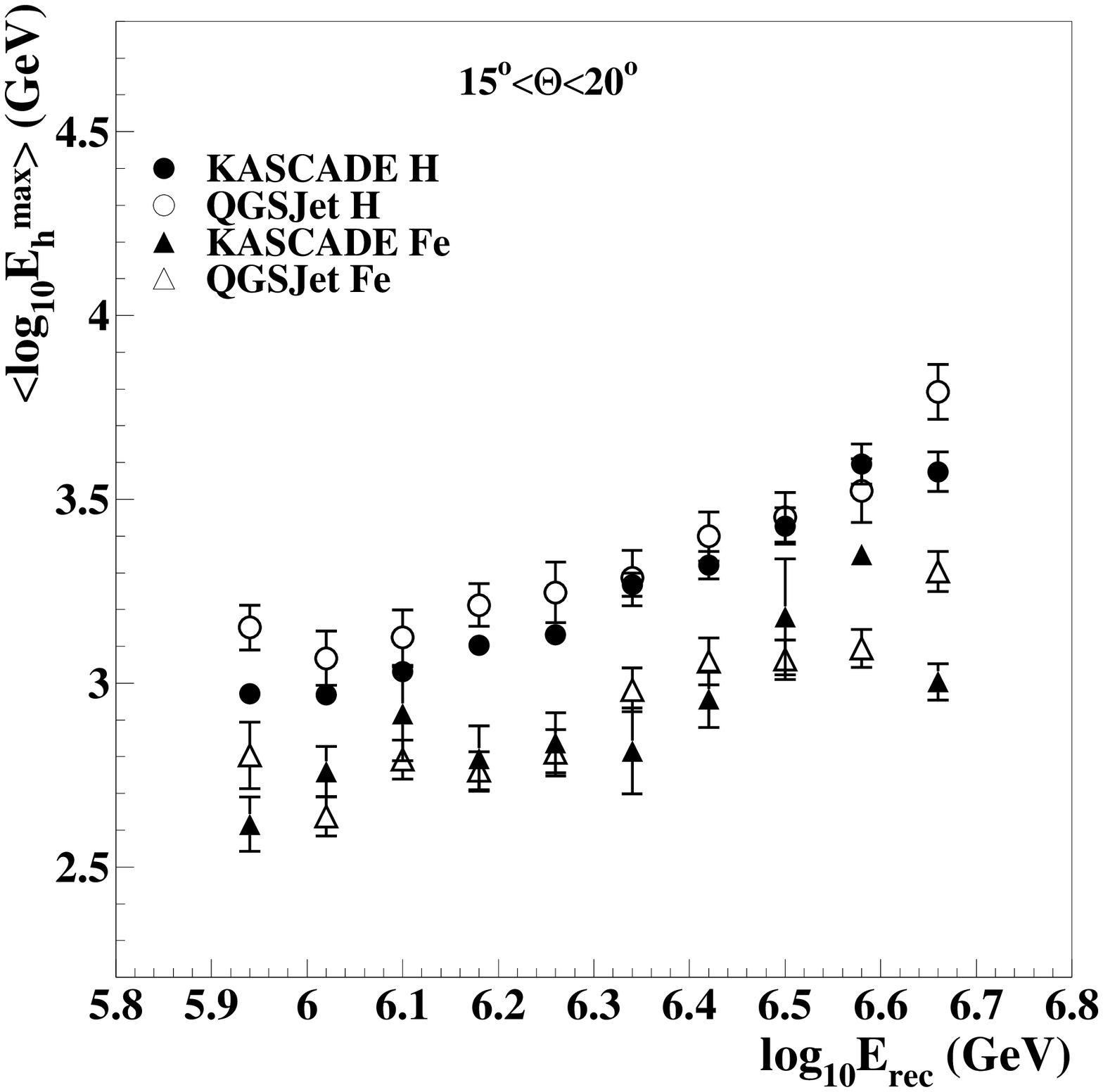,width=0.99\textwidth}
\vspace{-0.9cm}
\caption{\label{ehmax} \it{Dependence of the highest energy
reconstructed hadron on the reconstructed primary energy 
for simulated and experimental proton and iron classes.}} 
\end{minipage}
\vspace{0.5cm}
\end{figure}

\begin{figure}[ht]
\begin{center}
    \epsfig{file=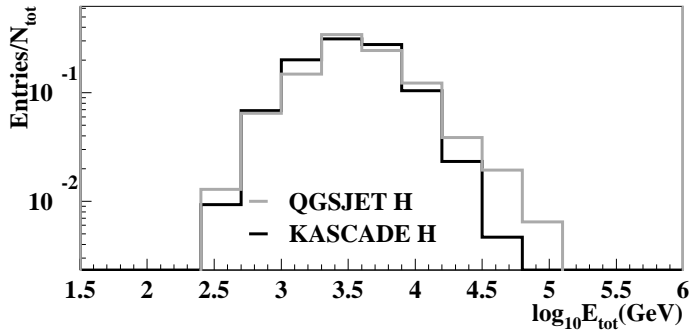,width=0.7\textwidth}
\end{center}
    \vspace{-0.2cm}
    \caption{\label{eh-test} \it One dimensional distributions of the 
    reconstructed hadronic energy sum for the proton enriched samples 
    of data and Monte Carlo in the energy range of 
    $6.05 < log(E_{rec}/GeV) < 6.25$.} 
\vspace{0.4cm}
\end{figure}

In Figure~\ref{eh}
we compare simulations using the QGSJET model~\cite{ostap1} with 
KASCADE data. 
In the data sample the fraction of showers 
induced by light nuclei is larger than that of heavy induced showers. 
Hence the distributions of the
proton class are smoother and show smaller
statistical fluctuations. 
For the data points with lowest reconstructed energies 
the model predicts an increased hadronic energy sum compared
to the data. 
This may be affected by systematic features of trigger or cut 
efficiencies, but the 
increase is even more pronounced if instead the total
energy sum the energy of the highest energy
reconstructed hadron is considered~(Figure~\ref{ehmax}).
This observation together with the information of earlier 
investigations of KASCADE data~\cite{roth01} suggests that cut and 
trigger effects are minor sources of the discrepancy.
The nonzero contamination of both, proton 
and iron induced events by the intermediate nuclei 
has been ignored in the simulations. 
This unknown role of primaries of intermediate mass
(mostly helium nuclei) at the data sample
may lead to the systematic differences in the distributions. 
Figure~\ref{eh-test} shows the one-dimensional 
distribution of the reconstructed hadronic energy for the proton
enriched sample in the energy range of 
$6.05 < log(E_{rec}/GeV) < 6.25$ for both data and Monte Carlo. 
Here again a slight overestimation of the hadronic energy at the 
predictions is visible. It was found that an adaption of
the simulated to measured slope of the energy spectrum 
does not reduce the deviation. But introducing a reasonable 
part of primary helium nuclei in the simulated sample lead 
to an overlap of the distributions. 

Allowing for the limited accuracy of the method, the distributions  
shown in Figures~\ref{eh}~and~\ref{ehmax}, 
originating from 
primary proton and iron nuclei do agree well with the 
predictions of the Monte Carlo simulations,
i.e. the overall dependence of the shower
observables are consistent with the simulations using the QGSJET
model. 
This finding is also confirmed by considerations of the numbers
of high-energy muons and hadrons reconstructed at the central 
detector.  
We therefore conclude that the QGSJET model 
describes the KASCADE 
data in a consistent way for the considered energy range 
of $10^{15}\,$eV - $6\cdot10^{15}\,$eV.  
The found behavior of the QGSJET model in the present 
investigations confirms the findings of an earlier publication 
of the KASCADE collaboration~\cite{hoerandel}. 
Especially the results shown in Figure~\ref{ehmax} are 
comparable to comparisons of Monte Carlo events with data 
in Figure~13 of ref.~\cite{hoerandel}, but in the present case 
the hadronic parameter is displayed with respect to the 
reconstructed primary energy and for enhanced samples of different 
mass groups.

With a larger sample of higher statistical accuracy this kind 
of comparisons provides, albeit indirectly, reliable information 
on strong interaction parameters and will help to tune
the hadronic interaction models used as Monte Carlo generators.
Studies with distributions based on larger statistical 
accuracy are foreseen for different energy intervals and various 
hadronic interaction models in a forthcoming paper.
A more sophisticated approach is planned with a neural network
trained for a classification in four or five mass groups. 

\section{Summary}
The paper presents an approach for the 
preparation of enriched cosmic ray mass group samples from EAS 
observations. For that purpose a unified framework of 
statistical inference has been used, based on nonparametric techniques 
for the analysis of multivariate parameter distributions. 
The approach has been demonstrated with experimental data of the 
multidetector experiment KASCADE, whose large number of EAS 
observables, simultaneously measured for each event, enables a 
successful application and useful investigations of high 
energy hadronic interactions on an event-by-event basis.

For the preparation of samples enriched with events of a 
given class (mass of the primary particle) the 
($N_{e}$,$N_{\mu}^{tr}$)-correlation has been exploited as 
a potential mass and 
energy identifier, using reference patterns from Monte Carlo 
simulations with the QGSJET model as generator of the 
high-energy hadronic interaction. 
The classification and misclassification rates have been 
studied and a purification procedure of the samples has been introduced. 
Efficiency and purity of the procedure are scrutinized. 
It has been shown that the purity of the samples (fraction of 
true classified events in an actual sample allocated to a 
given class) can be noticeably improved without a drastic
reduction of the efficiency 
(defined as fraction of true classified events of the 
total number of events of a given class). 
It should be emphasized that the procedure does remove events 
over nearly the full range of the distributions, thus avoiding 
any biasing of the remaining samples. In addition the approach 
accounts properly for the natural EAS fluctuations, as far as
the reference patterns reflect also these fluctuations
realistically. 
  
For illustrations, in a second step, the prepared samples 
have been used to study various EAS observables 
from the KASCADE experiment and their consistency with the 
QGSJET model. 
The electron size and muon content can be fairly 
well reproduced, as well as hadronic observables measured at the
core of showers, 
though in the latter case the samples are affected by the 
limited statistical accuracy of the number of events registered 
with the hadron calorimeter, and of the Monte Carlo simulations. 
An improvement of the analyses in this direction is a future task 
with studies of other current hadronic interaction models using
the presented distributions.

It is obvious that the demonstrated approach enables a number 
of interesting investigations of the interaction of such 
enriched samples with the air nuclei, e.g. studies of the 
attenuation lengths of specified primary cosmic particles by 
EAS observations with different zenith angles of incidence and 
with detector installations on different observation levels.

\section*{Acknowledgment}
{\it The authors would like to thank 
the members of the
engineering and technical staff of KASCADE who considerably
contributed to the reported measurements.
The work has been supported by the Ministry for Research of the 
Federal Republic of Germany and by a NATO Collaborative Linkage Grant
between Armenia and Germany (PST.CLG.978825).
The Romanian Ministry of Education and Research provided a grant for
supporting the collaborating group of the National Institute for
Physics and Nuclear Engineering of Bucharest. The Polish
collaborating group of the Cosmic Ray
Division of the Soltan Institute of Nuclear Studies in Lodz 
is supported by the Polish
State Committee for Scientific Research (grant No.~5 P03B 133 20).
The KASCADE collaboration 
work was embedded in the frame of scientific-technical co-operation 
(WTZ) projects between Germany and Armenia (ARM 98/002),
Poland (POL-99/005), and Romania (ROM 99/005).}   
